%% file: main.tex
\documentclass[compsoc,conference,a4paper,10pt,times]{IEEEtran}
\IEEEoverridecommandlockouts
\usepackage{cite}
\usepackage{amsmath,amssymb,amsfonts}
\usepackage{algorithmic}
\usepackage{graphicx}
\usepackage{textcomp}
\usepackage{bmpsize}
\usepackage{xcolor}
\usepackage{lipsum}
\usepackage{url}
\usepackage{balance}

\usepackage[colorlinks=true,urlcolor=black]{hyperref}
\def\BibTeX{{\rm B\kern-.05em{\sc i\kern-.025em b}\kern-.08em
    T\kern-.1667em\lower.7ex\hbox{E}\kern-.125emX}}

\begin{document}

\title{Supply Chain Insecurity: The Lack of Integrity Protection in SBOM Solutions}

 \author{\IEEEauthorblockN{Can \"{O}zkan}
 \IEEEauthorblockA{\textit{KU Leuven - COSIC} \\
 Kasteelpark Arenberg 10\\
 3001 Heverlee, Belgium \\
 can.ozkan@kuleuven.be}
 \and
 \IEEEauthorblockN{Xinhai Zou}
 \IEEEauthorblockA{\textit{KU Leuven - COSIC} \\
 Kasteelpark Arenberg 10\\
 3001 Heverlee, Belgium \\
 xinhai.zou@kuleuven.be}
 \and
 \IEEEauthorblockN{Dave Singel\'{e}e}
 \IEEEauthorblockA{\textit{KU Leuven - COSIC} \\
 Kasteelpark Arenberg 10\\
 3001 Heverlee, Belgium \\
 dave.singelee@kuleuven.be}
}

\maketitle

\begin{abstract}
The SolarWinds attack, which exploited weaknesses in a software update mechanism, highlights the critical need for organizations to have better visibility into their software dependencies and potential vulnerabilities associated with them. The Software Bill of Materials (SBOM) is paramount in ensuring software supply chain security. Under the Executive Order issued by President Biden, the adoption of the SBOM has become obligatory within the United States. The executive order mandates that an SBOM must be provided for all software purchased by federal agencies. In this paper, we present an in-depth and systematic investigation of the trust that can be put into the output of SBOMs. Our research reveals that the SBOM generation process across popular programming languages is susceptible to stealthy manipulation by malicious insiders, leading to significant supply chain insecurities. We then investigated the tools used to consume SBOMs, examining their capability to detect and handle manipulated or compromised SBOM data. To address these security issues, we analyze the use of public repositories for software libraries to validate the integrity of dependencies and demonstrate the feasibility of our proof-of-concept implementation. We further evaluate an alternative, decentralized approach based on blockchain.
\end{abstract}

\maketitle

\input{introduction}


\input{threat_model}

\input{malicious_developer_attack}

\input{sbom_consumption}

\input{solution}

\input{distributed}

\input{related_work}

\input{conclusion}

\input{acknowledgements}

\bibliographystyle{IEEEtran}
\bibliography{references.bib}
\balance
\end{document}

%% file: introduction.tex
\section{Introduction}\label{introduction}
\subsection{SBOMs vs supply chain security}

The software supply chain encompasses all processes involved in developing, delivering, and maintaining software, including implementing code, integrating and importing third-party libraries, building software artifacts, and distributing them to end users. Given the complexity and interconnectedness of modern software ecosystems, securing software supply chains has become paramount, especially in response to high-profile incidents such as SolarWinds~\cite{alkhadra2021solar}, CodeCov~\cite{CodeCovIncident}, and event-stream~\cite{EventStreamIncident}.

SBOMs have emerged as a crucial instrument and defense mechanism against software supply chain attacks. An SBOM is a formal, machine-readable inventory of software components and dependencies, information about those components, and their relationships~\cite{SbomGlance}. It provides an inventory of the various software components used in an application, such as libraries, frameworks, and dependencies. This nested inventory identifies the root and nested components within the software, allowing for better management and mitigation of potential security risks~\cite{USCommerce}. The adoption of SBOMs became vital after the Biden administration signed the Executive Order on Improving the Nation's Cybersecurity in May 2021~\cite{EO}. This order mentions that commercial software development often lacks transparency and adequate controls to prevent tampering by malicious actors, and therefore mandates the use of SBOMs across government agencies and critical infrastructure sectors.  For this reason, any company that sells software to the federal government must provide the application and a corresponding SBOM~\cite{sonatype}.

An SBOM encompasses various formats for documenting and sharing information about software components within the supply chain. From an abstract point of view, there are two distinct phases in the lifetime of an SBOM: (i) the generation of the SBOM (typically at the supplier side) and (ii) the use/consumption of the SBOM (typically at the end user's side). Two prominent SBOM types, CycloneDX~\cite{cyclonedx} and SPDX~\cite{spdx} (Software Package Data Exchange), have emerged as leading standards in this domain. The main use cases of SBOMs are license and vulnerability management. In this paper, we focus on the latter. SBOMs can facilitate vulnerability management as follows. By cross-checking the output of an SBOM and a vulnerability database such as the National Vulnerability
Database (NVD), any end user having the SBOM can identify known vulnerabilities in the software components of a product, and (ideally) address these -- for example by patching. This process can then be repeated at regular intervals. 

Since SBOMs will be at the direct basis of the vulnerability management process, the integrity of the information provided in the SBOM is paramount. Let us for instance consider the following example. The Poco library for C++~\cite{pocolib} has a buffer overflow vulnerability for versions prior to 1.13. If a software project uses an old version this library, e.g., the 1.9 version, the SBOM file should show this so that this can be fixed. However, if an attacker would succeed in manipulating the SBOM generation process such that the SBOM file would show version 1.13, the vulnerability within the Poco libarary in the software project will remain hidden for the end user. In summary, such manipulations can lead to misleading vulnerability reports being generated by SBOM consumption tools, resulting in organizations inadvertently believing that their software assets are free from vulnerabilities. 

\subsection{Improving supply chain security}
\begin{figure}
    \centering
    \includegraphics[width=\linewidth]{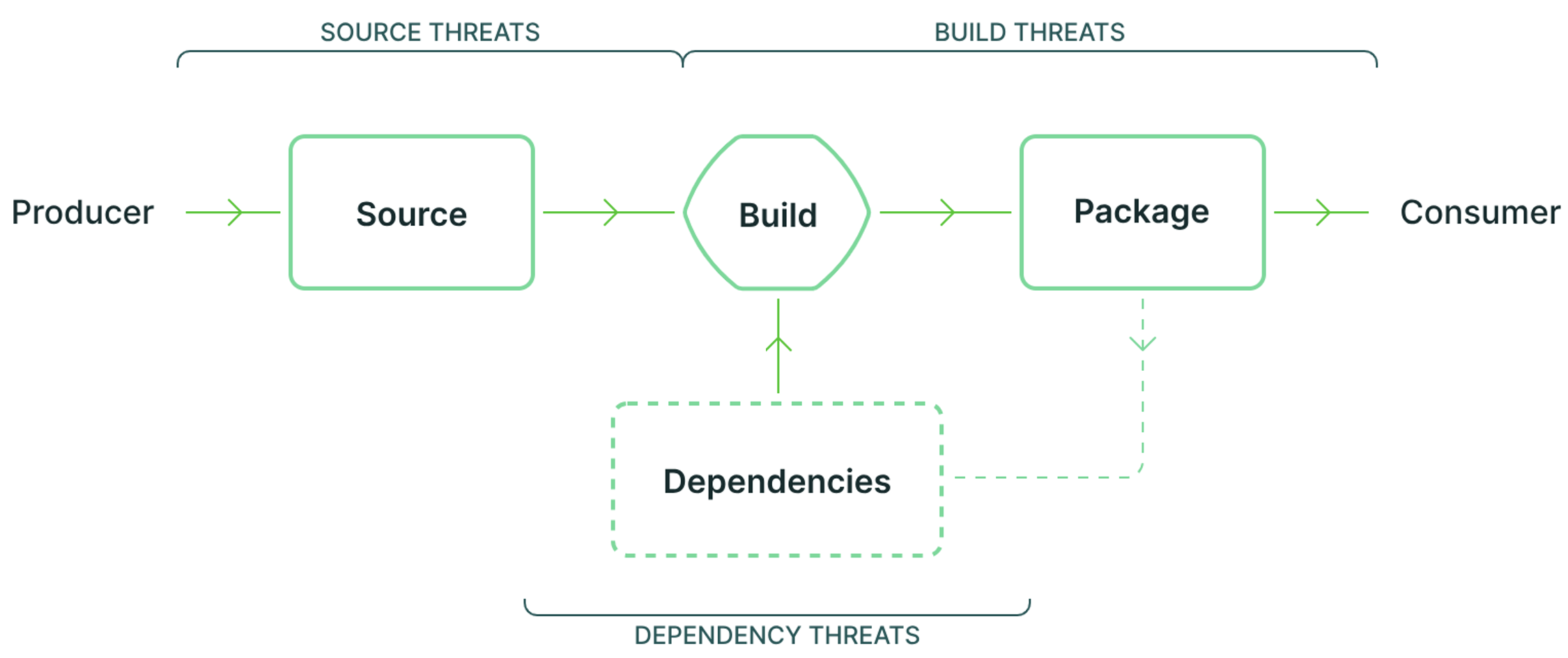}
    \caption{SLSA Threat Model}
    \label{fig:slsa}
\end{figure}

A great deal of effort has been exerted to improve supply chain security. The SLSA~\cite{SLSA} framework was introduced by Google to investigate attack entry points in the software supply chain. Given the importance of this model, we use this as a main reference in our threat model in Section~\ref{ThreatModel}. From a high-level point of view, the framework categorizes software supply chain threats into three distinct categories, as depicted in Fig.~\ref{fig:slsa}: \textit{source threats}, \textit{built threats}, and \textit{dependency threats}. Source threat involve unauthorized changes to the original source code or code base. Built threats occur when adversaries infiltrate or manipulate the built process or infrastructure. Lastly, dependency threats arise when attackers compromise external libraries or packages integrated into the software, which propagates harmful behavior into dependent projects. 

On the defensive side, \textit{In-toto}~\cite{torres2019toto} is a framework designed to secure the software supply chain by tracking and verifying each step of software development, from code creation to the final distribution of software artifacts. It uses cryptographic verification to generate provenance metadata, which allows consumers to validate that the software was built exactly as intended without unauthorized modifications. Furthermore, \textit{cosign}~\cite{newman2022sigstore} provides a simplified approach and digitally signs and verifies container images and software artifacts. It helps ensure the integrity and authenticity of software by enabling users to detect any unauthorized or malicious changes before deployment or consumption by the end user. 

However, there are specific scenarios where these tools may not suffice. In-toto utilizes artifact rules to define three core elements for each step in the supply chain: expected inputs, expected outputs, and the specific relationships between them (allow, disallow, require, create, delete, modify, and match). For instance, an artifact rule may state that source code files must originate exclusively from a specific repository and match a certain cryptographic hash, ensuring they have not been tampered with. Package managers typically utilize configuration files, such as Python's requirements.txt or the Node.js's package.json, to define dependencies, their versions, and related configuration details. These configuration files are essential for managing project dependencies and often need regular updates or modifications, for example, when adding new libraries or adjusting versions for compatibility or a vulnerability fix on the supplier side. Therefore, in frameworks like in-toto the artifact rules need to explicitly allow legitimate modifications to such files. However, a direct consequence is that this opens up entry points for an attacker to introduce vulnerabilities, as we will show later.  Similarly, software and the corresponding SBOM are generated on the supplier side; then, those artifacts are signed to be distributed. Therefore, cosign does not protect against modifications made on the supplier side. In cosign, it is also crucial to clearly define which artifacts need signing, when they should be signed, and whether signing should be limited to specific files or extended more broadly to all software artifacts.


\subsection{Contributions and outline of the paper}
In this paper, our contribution is four-fold: 
\begin{enumerate}
    \item We show that for most programming languages, a malicious insider can manipulate the SBOM generation process in a stealthy way, even when supply chain security solutions such as in toto are in place. 
    \item We perform an analysis of popular SBOM consumption tools and identify several security weaknesses.
    \item We show that the above-mentioned insider attack can be mitigated by relying on central public repositories for software libraries.
    \item We implement a proof-of-concept for the Python ecosystem to demonstrate the feasibility of our proposed solution.
    \item We evaluate an alternative design approach for a decentralized setting.
\end{enumerate}

The rest of the paper is organized as follows: In Section \ref{ThreatModel}, we describe our assumptions, threat model, and the entry point for the attack on the software supply chain ecosystem. In Section \ref{malicious_developer_attack}, we zoom in on the malicious insider attack and discuss how SBOM generation tools can be mislead in outputting incorrect dependency information and version numbers. In Section \ref{sbom_consumption}, we extend our security study and analyze the security of existing SBOM consumption tools. Section \ref{solution} presents our mitigation strategy based on central public repositories for software libraries. In Section \ref{decentralized}, we investigate an alternative, decentralized approach. Finally we give an overview of related work in Section \ref{Relatedwork} and conclude the paper in Section \ref{conclusion}.   



%% file: threat_model.tex
\section{Threat Model}\label{ThreatModel}

The Software Development Life Cycle (SDLC) is a framework that manages the software development process and breaks it into planning, development, testing, deployment, and maintenance phases. However, modern software development heavily relies on third-party components, open-source libraries, and automated build pipelines, intrinsically linking SDLC to the software supply chain. The software supply chain includes all external and internal dependencies in software development, from third-party libraries to CI/CD pipelines and software distribution mechanisms.

One can identify three high-level attack points during the SDLC: (i) during the development phase at the supplier; (ii) during distribution of the software; and (iii) when used by the end-user. This high-level categorization aligns with the SBOM life cycle as there are three distinct phases in SBOM: generation, distribution, and consumption. Each phase can individually become a target for attackers attempting to compromise the integrity or authenticity of software artifacts, such as an SBOM file. 

In our paper, we assume that after the generation of the SBOM, it is securely signed by the supplier. Moreover, we assume that the online distribution of software and its artifacts (such as an SBOM) are protecting used TLS. Later in this paper (Section~\ref{sbom_consumption}), we revisit some of these assumptions and show that these often do not hold in practice. However, for the sake of simplicity, let us for now assume that the distribution and consumption phase of the SBOM is properly secured. We also assume that other supply chain mechanisms such as in-toto and cosign are in place and correctly configured. Given these assumptions, we zoom in on the SBOM generation phase. 

Recall that our main focus is supply chain security and that security mechanisms such as in-toto are in place. Therefore we consider an attacker that is authorized to make changes to the software and wants to introduce potential attack points because of the use of insecure software components (e.g., older and insecure versions of a particular software library). Clearly the attacker wants to hide these security vulnerabilities in the SBOM. Otherwise any SBOM consumer would trivially notice these vulnerabilities and would patch these (or just not use the software). Given this attack setting, let us now revisit the SLSA framework depicted in Fig.~\ref{fig:slsa} and further zoom in on the attack points during the software development phase. The SLSA framework categorizes software supply chain threats into three key areas: source threats, dependency threats, and built threats. From these three categories, the source threats are the most relevant one. Indeed, during this phase of the SDLC, one is authorized to make changes to the software code base. Therefore tools such as in-toto will allow these changes to take place. 

\begin{figure}
    \centering
    \includegraphics[width=\linewidth]{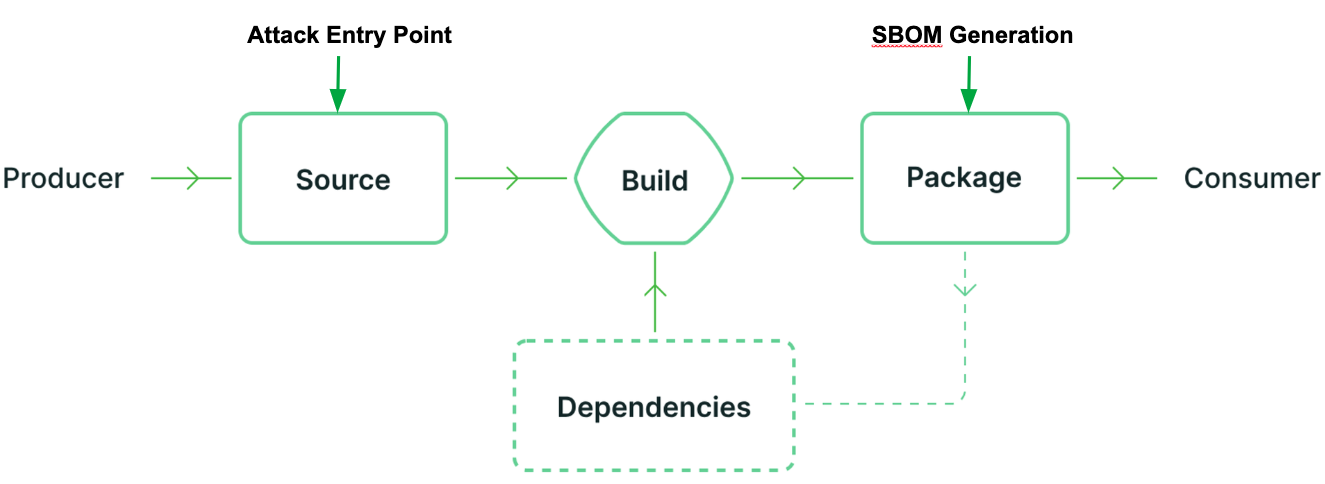}
    \caption{Attack Entry Point vs SBOM generation}
    \label{fig:slsa2}
\end{figure}

In practice, we could see two potential attack scenarios for the source threats. The first one is a malicious insider, for example disgruntled developer or contractor, who intentionally introduces harmful changes directly into the repository and influences the SBOM generation process in such a way that the harmful changes remain invisible when the SBOM is generated later during the software development process. The second attack scenario is a compromised machine of the developer that introduces these vulnerabilities to the code. This could for example be the result a specific exploit executed on that machine or the developer being victim of phishing. In both cases, malicious changes are made to the software and the SBOM generation process is manipulated so that the SBOM will be wrong. It is important to stress that the attacker (i.e., malicious insider) does not generate the SBOM, this is done by another entity in the software development chain. The SBOM is typically generated at the end of the packaging phase, when the software is ready for distribution, as can be seen in Fig.~\ref{fig:slsa2}.

%% file: malicious_developer_attack.tex
\section{Malicious Developer Attack}\label{malicious_developer_attack}

The SBOM generation process entails creating a list of all software components, such as libraries, frameworks, and dependencies, and their corresponding versions used in an application. Available automated tools can scan the source code and file system in order to reveal dependencies, libraries, frameworks, and other components used in the application. In addition, many modern programming languages and frameworks utilize package and dependency management tools such as Maven~\cite{maven}, npm~\cite{npm}, pip~\cite{pip}, and composer~\cite{composer} that maintain a list of external dependencies imported and their versions so that those dependencies can be easily installed at the beginning of the software installation or when another developer forks the code base. SBOM generation tools, programs that help organizations produce their SBOM, typically retrieve information to identify and document dependencies through the aforementioned methods.

In this section, we now focus on the malicious insider attack mentioned before in the paper, and perform a comprehensive analysis and experimental evaluation of this attack across various prominent programming languages. Recall that the goal of the attack is to manipulate the SBOM generation process in a stealthy way, so that when an SBOM is generated at a later stage in the development cycle it will be incorrect. Manipulating the SBOM generation process to exclude specific dependencies or to show misleading version numbers compromises the integrity and accuracy of the resulting SBOM artifact. 

We conducted experiments to compromise the integrity of the SBOM generation process by manipulating dependency version information in package managers. More specifically, we analyzed the SBOM generation process for Python, C, C++, C\#, Java, Javascript, Rust, and PHP. These are the first eight programming languages based on Tiobe's November 2023 ranking~\cite{TiobeIndex}, and the Rust language has gained huge popularity over the years. As our research focus is security, we selected the tools for SBOM generation from the CycloneDx tool center. The tools we experimented with for this section are CycloneDx for Python~\cite{CyclonedxPython}, CycloneDx for Conan~\cite{CyclonedxConan}, CycloneDx for Maven~\cite{CyclonedxMaven}, CycloneDx for .NET~\cite{CyclonedxDotnet}, CycloneDx for npm~\cite{CyclonedxNodeNpm}, CycloneDx for PHP Composer~\cite{CyclonedxPHP}, Cyclonedx Rust Cargo\cite{rustsbom}, and Syft~\cite{syft}.

For each of these tools, we performed a source code review to reveal the SBOM generation process and how the tools identify the dependency information. This in-depth analysis aims to clarify the mechanisms through which these tools extract and process dependency information. Our analysis revealed that one can compromise the integrity of the generated SBOM data by manipulating the package managers and other files in all tools except CycloneDx for Maven and Cyclonedx Rust
Cargo. The generation process is susceptible to manipulation through tampering with package management systems and their associated dependency files. Dependency files such as package.json and requirements.txt specify the software project's external dependencies and versions. An adversary can easily introduce misleading information by tampering with these dependency files and additional files. We will discuss this more in detail in the following paragraphs. This finding underscores the importance of implementing robust security mechanisms within SBOM generation tools and that these should not solely rely on dependency and file information.

The next paragraphs dissect the SBOM generation tools specific to each programming language. The results are summarized in Table~\ref{tab:generationresults}.

\textbf{PHP:} It is possible to manipulate the SBOM generation process for PHP projects. Manipulating the contents of the \textbf{./composer.json}, \textbf{./composer.lock}, and \textbf{/vendor/composer/installed.json} files in the code base influence the SBOM generation process, thereby potentially compromising the reliability and accuracy of the resulting SBOM artifact. Another interesting outcome is that CycloneDx for PHP does not produce a hash value for components. Since a hash value is in practice not often utilized in the consumption phase, as will be discussed in the next section, it does not break the consumption operation. However it is essential information to guarantee the integrity of dependency information. \\

\begin{figure}
    \centering
    \includegraphics[width=\linewidth]{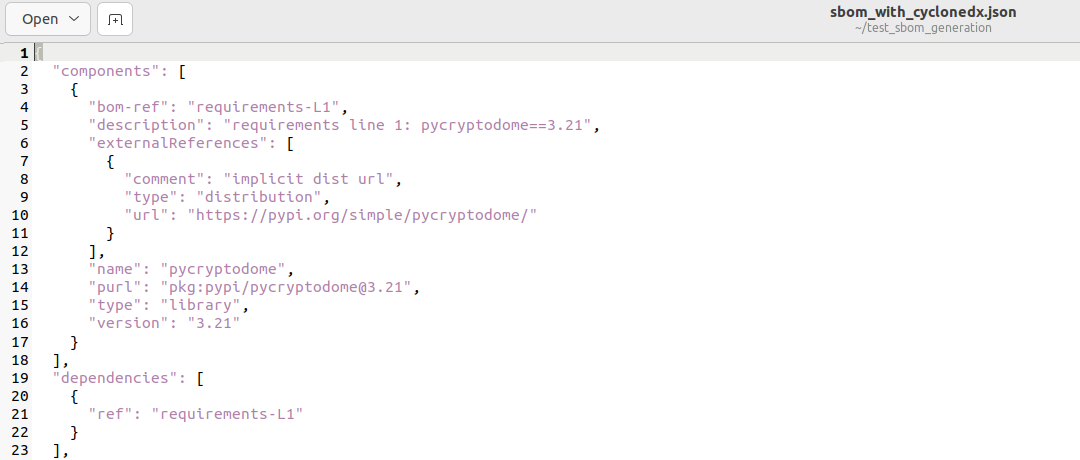}
    \caption{Cyclonedx-Py: SBOM with no hash value for dependencies}
    \label{fig:cyclonedxpy}
\end{figure}

\textbf{Python:} It is possible to manipulate the SBOM generation process for projects written in Python. The requirements.txt file is a package management file and dictates the dependencies required for the project's execution. The SBOM generation process can be manipulated by changing the contents of the \textbf{requirements.txt} file in the code base. Another interesting observation is that CycloneDx for Python produces no component hash values, as can be seen in Fig.~\ref{fig:cyclonedxpy}.  \\

\begin{figure}
    \centering
    \includegraphics[width=\linewidth]{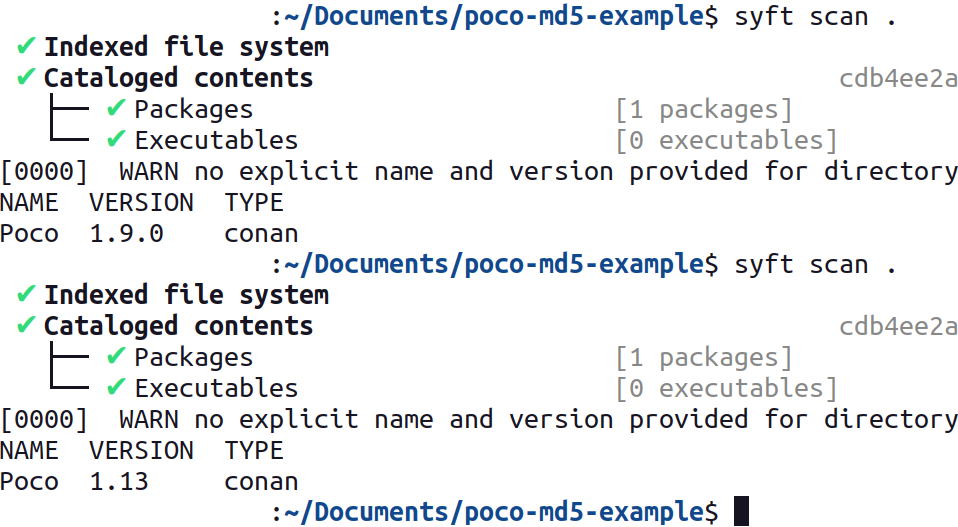}
    \caption{Manipulating software versions in C/C++}
    \label{fig:ccplus}
\end{figure}

\textbf{C/C++:} The tool used to generate SBOM for C/C++ projects is read-only at the time of writing~\cite{CyclonedxConan}. The owner archived this repository on Oct 2, 2023. Although there is a newer version of the tool ~\cite{CyclonedxConan2}, it is not yet a mature project to generate SBOM for C/C++ projects. Hence, we could not analyze the CycloneDX Conan SBOM generation tool. Instead, we used another tool. Syft, supporting C/C++, Debian, Go, and JavaScript ecosystems, among others, was used to generate an SBOM for C/C++ projects. It supports the Conan package manager for C/C++ projects. It is possible to manipulate the SBOM generation process for C/C++ projects utilizing the Conan package manager by an adversary manipulating either \textbf{conanfile.txt} or \textbf{conanfile.py}. We show an example in Fig.~\ref{fig:ccplus}. In this example, we changed the version of the Poco library from 1.9 (vulnerable to buffer overflow) to 1.13 (not vulnerable), so that the SBOM will show the incorrect version. \\

\begin{figure}
    \centering
    \includegraphics[width=\linewidth]{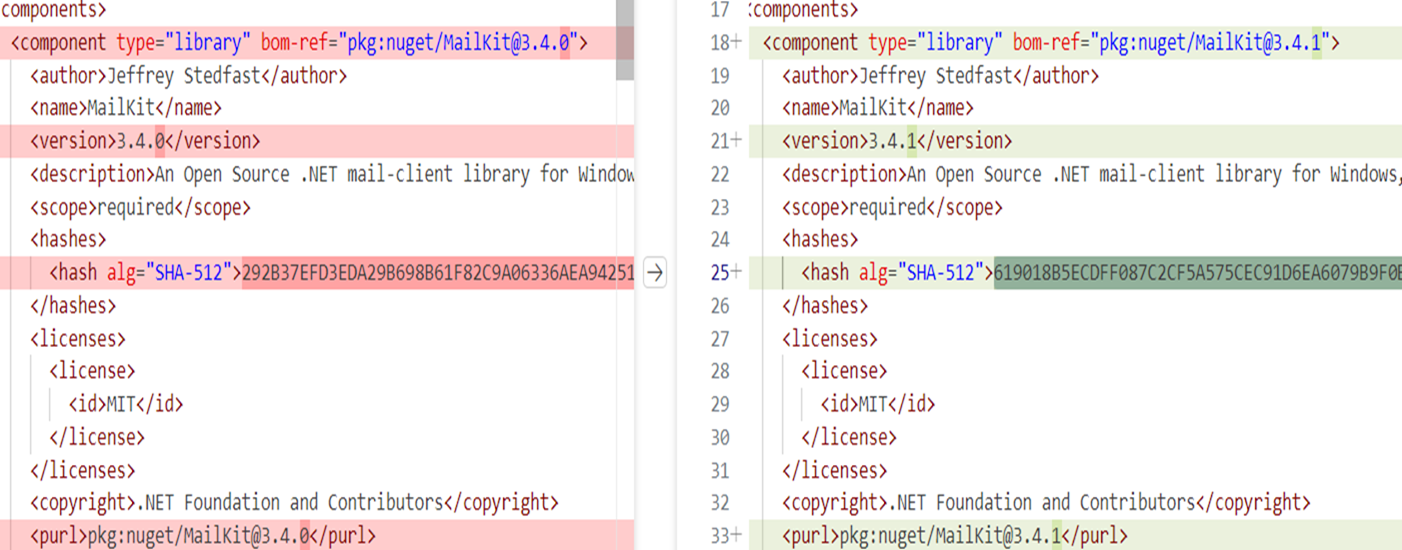}
    \caption{Different Hash Values after manipulating version numbers in C\#}
    \label{dotnt}
\end{figure}

\textbf{C\#:} It is possible to manipulate the SBOM generation process for dotnet projects written in C\#. The tool used to generate the SBOM focuses on dotnet projects written in C\#. Changes made to the \textbf{packages.config} file in the code base affect the SBOM generation process, which could make the final SBOM artifact incorrect. Another observation is that the SBOM generation tool produces different hash values once you alter version numbers, as shown in Fig.~\ref{dotnt}. \\

\textbf{Java:} It is not possible to tamper with Java projects that utilize Maven for their build process. There are two tools to generate an SBOM for Java projects: CycloneDX Maven Plugin~\cite{CyclonedxMaven} and CycloneDX Core (Java)~\cite{CyclonedxJava}. Both are used as Maven plugins. Hence, we focused on the CycloneDX Maven Plugin. The plugin configuration for usage is available in ~\cite{MavenUsage}. Maven, a widely used build and package management tool in the Java development ecosystem, has a technique for efficiently managing dependencies. Maven prominently includes a centralized repository that is a reliable and authoritative source for pre-compiled software libraries and dependencies. When a Java project uses Maven, all changes made to the project object model (pom.xml) file swiftly initiate an immediate and efficient updating process. This automated process guarantees that the project includes the designated versions of the dependencies listed in the pom.xml file by retrieving them from the central repository. If an adversary tampers with the version numbers in pom.xml, Maven will download and install the versions the attacker indicated. For this reason, an adversary trying to sabotage the SBOM generation process (i.e., hiding insecure software libraries) might actually patch the vulnerability in a dependency while trying to manipulate the process. Thus, the SBOM generation process for Java projects leveraging Maven for build and dependency management does not allow tampering with the version numbers. \\

\begin{figure}
    \centering
    \includegraphics[width=\linewidth]{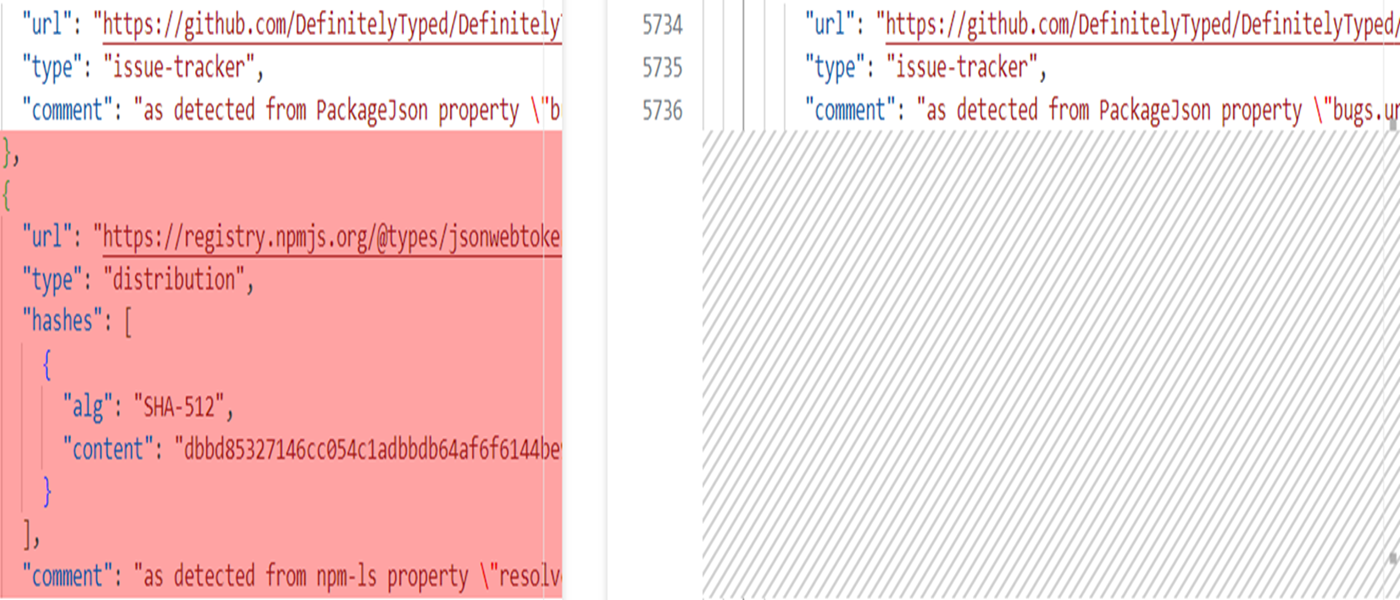}
    \caption{No Hash Values shown when tampering with dependencies in JS}
    \label{fig:jsnohash}
\end{figure}

\textbf{JavaScript:} It is possible to manipulate the SBOM generation process for Node.js projects, an open-source cross-platform JavaScript runtime environment written in JavaScript, if adversaries manipulate the package.json, package-lock.json, and package.json files under the node\_modules folder. Furthermore, we observed that the SBOM generation tool generates a hash value when the generation process was not tampered with. In contrast, it does not generate a hash value for any component in a software project when the process has been manipulated, even if only one version number has been tampered with. In other words, only one alteration of a version number is enough to not generate any hash value for all the components in a software project, as seen in Fig.~\ref{fig:jsnohash}. \\

\begin{figure}
    \centering
    \includegraphics[width=\linewidth]{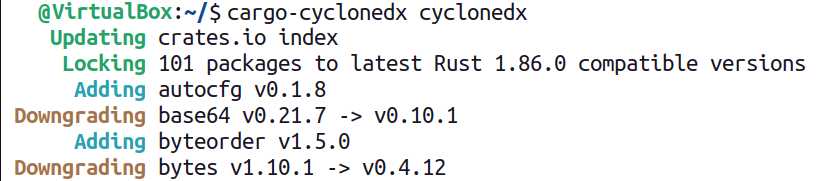}
    \caption{Rust Immediate Update Process}
    \label{fig:rust}
\end{figure}

\textbf{Rust:}Rust uses Cargo for its package manager. It is not possible to tamper with Rust projects. Generating an SBOM using cyclonedx-rust-cargo swiftly initiates an immediate and efficient updating process, as seen in Figure \ref{fig:rust}. This automated process guarantees that the project includes the designated versions of the dependencies listed in the Cargo.toml file by retrieving them from the central repository.

\begin{table*}
    \centering
    \caption{SBOM Generation Tampering Results}
    \label{tab:generationresults}
        \begin{tabular}{|c|c|c|}
            \hline
            \textbf{Programming Language} & \textbf{Tampering Possible} & \textbf{Files that need to be manipulated} \\ \hline
            Python & X & requirements.txt \\
            C & V & conanfile.txt, conanfile.py \\
            C++ & V & conanfile.txt, conanfile.py \\
            C\# & V & packages.config \\
            Java & X & N/A \\
            JavaScript & V & package.json, package-lock.json, and related dependency folders under node\_modules \\
            PHP & V & composer.json, composer.lock, /vendor/composer/installed.json \\
            Rust & X & N/A \\
            \hline
        \end{tabular}%
\end{table*}

From the analysis above, one can conclude that for 6 out of the 8 programming languages investigated, it is possible to manipulate the SBOM generation process in a stealthy way. During our analysis, we also made another security-related observation. While digital signatures can guarantee the authenticity and integrity of SBOM data, none of the SBOM generation tools had their used activated by default ~\cite{muiri2019framing}. Since their use is optional, there is a large risk that these are not used in practice, allowing adversaries to trivially manipulate an SBOM.

We further also observed that CycloneDx for Python, CycloneDx for PHP Composer, and Syft tools do not generate hash values for dependencies to be included in an SBOM file, whereas CycloneDx
for npm, Cyclonedx for dotnet, and Cyclonedx for Maven generate hash values. As will be discussed later in Section \ref{solution}, having a hash value for each dependency is an essential component when aiming to mitigate stealthy malicious insider attacks.

%% file: sbom_consumption.tex
\section{Analysis of existing SBOM consumption tools}\label{sbom_consumption}
Before we discuss how to mitigate the malicious insider attack, we first want to focus on the SBOM consumption process. Up to now, we assumed that the consumption of an SBOM was done securely. In this section, we aim to assess whether this assumption holds in practice. Therefore we analyzed existing SBOM consumption tools. We systematically reviewed four distinct consumption applications and analyzed how these tools react when the version numbers in package managers are tampered with. Again as before, we chose the tools recommended in the CycloneDx tool center. The SBOM consumption tools we experimented with are the OWASP Dependency Track ~\cite{dependencytrack}, Bomber ~\cite{bomber}, cve-bin-tool ~\cite{cvebintool}, and Grype ~\cite{grype}. The criteria for tool selections stem from their adherence to fundamental principles such as open-source availability, unrestricted free accessibility, and formal recognition by the CycloneDX standard. Our methodology consists of a systematic source code review followed by a proof-of-concept (PoC) validation. Initially, we conducted an analysis of the source code of available tools to understand how consumption tools generate vulnerability data for a given SBOM file. 

Our analysis revealed a security issue within all four applications. None of the applications employed any cryptographic controls for integrity verification in the dependencies listed in the SBOM file. If hash values were present within the SBOM file, they were not checked to ensure the integrity of the libraries and dependencies or to detect potential manipulation. Instead, these tools rely solely on the version number shown in the SBOM to identify possible vulnerabilities, which is clearly a shortcoming from a security point of view.

\begin{figure}
    \centering
    \includegraphics[width=\linewidth]{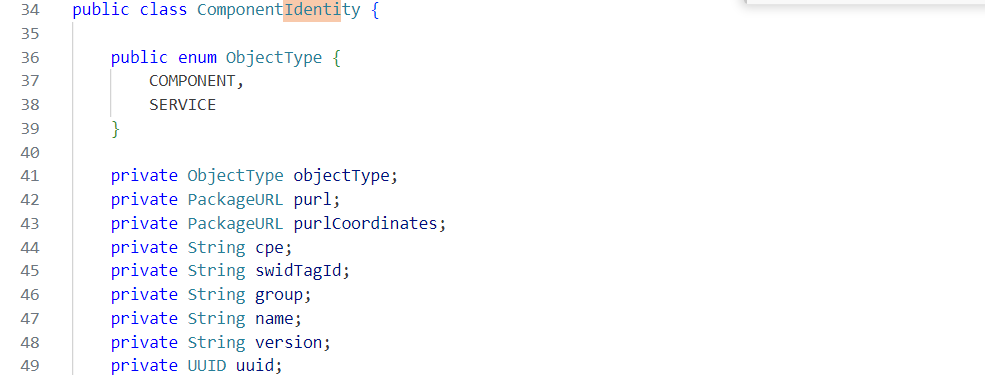}
    \caption{Component information}
    \label{fig:componentinfo}
\end{figure}

\begin{figure}
    \centering
    \includegraphics[width=\linewidth]{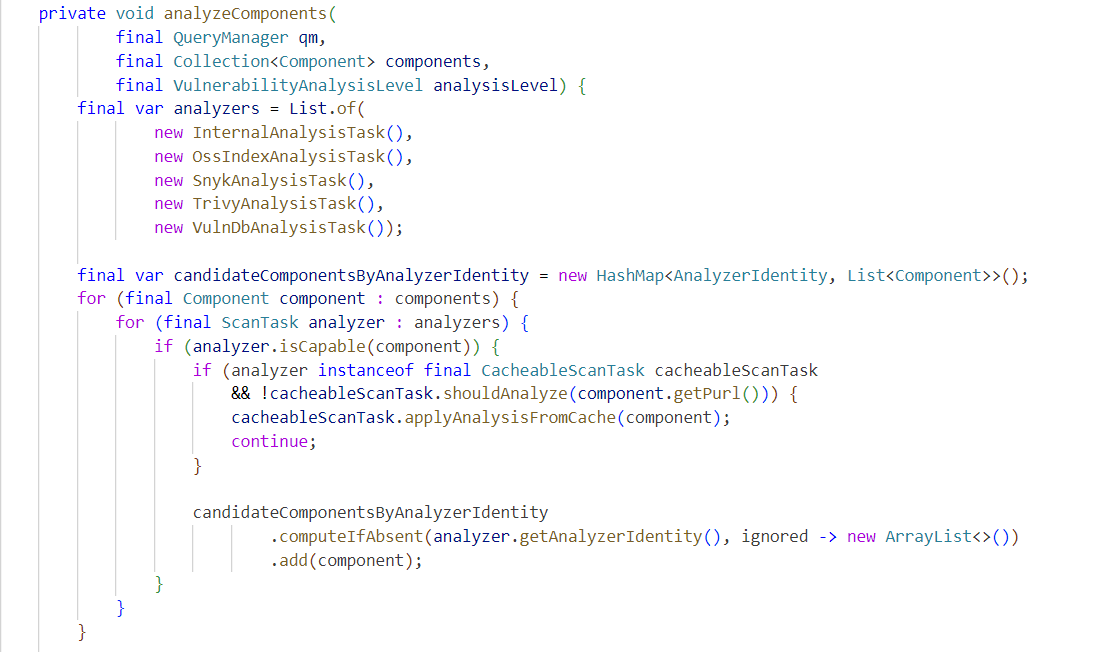}
    \caption{Vulnerability Analysis Task}
    \label{fig:vulnerabilityanalysis}
\end{figure}

We will now examine Dependency-Track in-depth, particularly focusing on its mechanism for identifying vulnerabilities and analyzing software components. Analyzing its source code reveals a structured process for identifying vulnerabilities in software components. The key steps include (i) component identification and version extraction and (ii) vulnerability analysis. Upon uploading an SBOM to Dependency-Track, the platform parses the SBOM to extract details about each component, including its name, version, and package URL (PURL), as depicted in Figure \ref{fig:componentinfo}. This parsing is managed by the BomUploadProcessingTask class, which processes the SBOM and registers the components in the system. Then, the VulnerabilityAnalysisTask class initiates the vulnerability analysis process. This task cross-references the identified components against various vulnerability databases to detect known vulnerabilities associated with the specific component versions, as seen in Figure \ref{fig:vulnerabilityanalysis}. Thus, it does not provide any hash control to verify the integrity of each component in a project in the vulnerability identification phase.

Another shortcoming we identified, was that the tools in question do not require SBOMs to be digitally signed before consumption, which presents a security risk. Consequently, even when suppliers digitally sign an SBOM file to confirm its integrity and authenticity, adversaries could effortlessly tamper with it and remove the signature before consumption (since this is not checked). It is evident that the lack of verifying digital signatures clearly opens the doors for trivial manipulation attacks on SBOMs. 

To experimentally validate our findings, we implemented a proof-of-concept attack where we tampered with the version numbers in the SBOM (so basically just changing the SBOM data) and then feed this manipulated SBOM file to one of the four SBOM consumption tools. We then compare the vulnerabilities reported by the tools to what is reported when a non-manipulated SBOM is consumed. Our experiment confirmed that this trivial manipulation of SBOM data is sufficient to have the tools report incorrectly about the vulnerabilities in the software. 

When performing these experiments, we observed another shortcoming. We noticed that each tool relied on a distinct vulnerability database, leading to discrepancies in the number of vulnerabilities identified by the SBOM consumption tools. OWASP Dependency Tracker utilizes the NVD and the GitHub Advisory database as primary sources ~\cite{dependencytracksource}. On the other hand, Bomber uses the Open Source Vulnerabilities (OSV) database ~\cite{bombersource} by default, and the cve-bin tool uses vulnerability information from NVD, OSV, Gitlab Advisory Database (GAT), RedHat Security Database (REDHAT), and Curl Database (Curl) sources ~\cite{cvebintoolsource}.

\begin{table}
        \centering
        \caption{Summary of Consumption Tools}
        \label{tab:table1}
        \begin{tabular}{|c|c|c|}
            \hline
            \textbf{Tool} & \textbf{Tampering} & \textbf{Signing Mandatory}  \\ \hline
            Dependency-Track&  V &  X \\
            Grype&  V &  X \\
            cve-bin-tool&  V &  X \\
            bomber&  V &  X \\
                          
            \hline
        \end{tabular}
    \end{table}

The conclusion of our analysis of existing SBOM consumption tools is that future versions of these tools should mandatory enforce digital signatures before consuming any SBOM, and ideally also verify hash values (see next section).

%% file: solution.tex
\section{Our proposed solution}\label{solution}

\subsection{Main concept}
\subsubsection{Linking SBOMs to the actual software}
As shown in Section~\ref{malicious_developer_attack}, the root cause of the malicious insider attack is that there is no strong link between an SBOM and the actual software. If such a link would exist, then would be able to detect that the version numbers and dependencies shown in the SBOM would not represent the actual software in case of manipulation attacks. In theory, the disconnection between software and the corresponding SBOM could occur during either SBOM generation, SBOM consumption, distribution and storage. Since the latter two stages can be trivially solved by using state-of-the-art security measures, we do not focus on these in our paper. In Section~\ref{sbom_consumption}, we demonstrated that popular SBOM consumption tools have security weaknesses. However, these can be again mitigated by using appropriate security solutions such as mandatory digital signature verification. As a result, the only remaining phase where the connection between the software and the SBOM can be broken, is during the SBOM generation phase. 

 Currently, SBOM tools use version names instead of hash values to represent software. Therefore, there is no strong connection between the version name and the actual software component (i.e., the dependency itself). The way forward is to compute the hash of the software and tie this to the SBOM. This provides a strong link, because even changing one bit of code would result in a completely different hash value. Of course, putting hash values in an SBOM is not sufficient. Nothing prevents an attacker from manipulating a software library to compute the corresponding hash value (after the attack) as well. What one actually needs, is a reference repository for hash values of software libraries. Lets take the example from before to illustrate this. If a malicious insider attack would change the version number of the C++ Poco library shown in the SBOM from 1.9 to 1.13, one could verify whether the hash of the software library corresponds to the reference hash of version 1.13 of the C++ Poco library. If the latter would be stored in an immutable repository, then the result of this hash verification would be false in case of an attack.
 
 \subsubsection{Framework for a secure reference repository}
 It is not straightforward to establish such an immutable reference repository for hash values of software libraries. The challenges arise from the complexity of the current software environment. Firstly, the huge amount of software makes it difficult to record hash values for all of them, necessitating an efficient storage mechanism. Secondly, the system is dynamic: software is frequently updated, sometimes even with different authors/owners. Thirdly, the system must handle high throughput and fast response times, as it will receive numerous read and write requests in parallel. Finally, some companies may need a private system for software used exclusively within their internal networks. Optionally, the solution can be decentralized, as no single party can be trusted on a global scale to manage a reference repository. The combination of these challenges makes the creation of a reference repository for storing hash values of software libraries a difficult task. Currently, to the best of our knowledge, no repository exists that is specifically designed for SBOM tools to verify whether information such as hash values have been modified. As a result, there is no ground truth. 

\begin{figure}[htbp]
    \centering
    \includegraphics[scale=0.3]{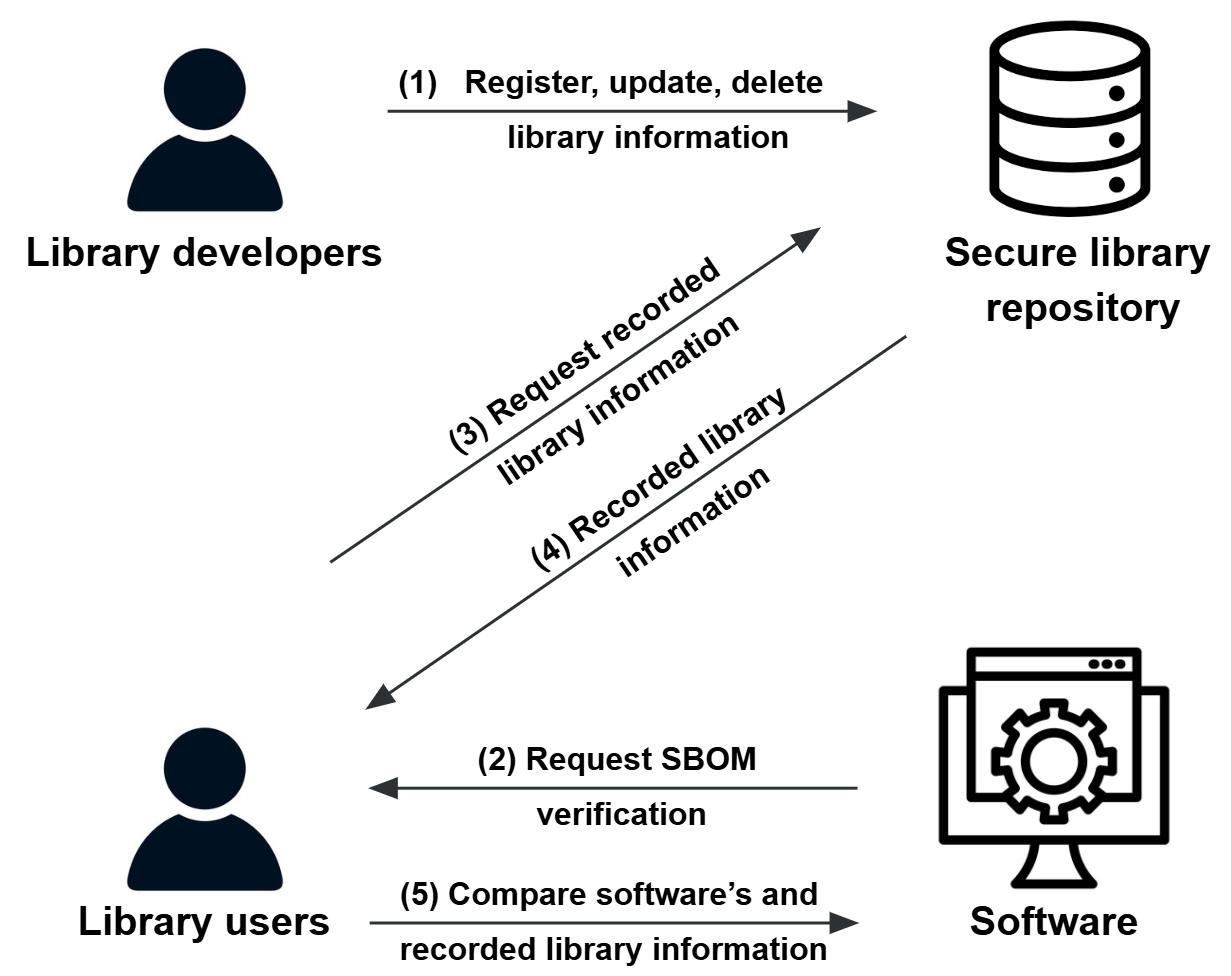}
    \caption{Proposed framework for the secure generation of SBOMs}
    \label{fig:sbom_generation}
\end{figure}

Our proposed framework for secure SBOM generation is illustrated in Fig.~\ref{fig:sbom_generation}. There are two primary stakeholders: software library developers and users. Developers are responsible for registering, updating, and deleting their software library information from the repository, while users focus on verifying the libraries of the software they are intending to use. The repository must securely store both all information to the software library as well as the identities of the software library developer (to claim ownership of the software and being authorized to make changes to the information stored). The process begins when (1) developers register or update their libraries. Library users can then (2) request SBOM verification by (3) making a query to the repository. Once users (4) retrieve the necessary information about the software library (e.g., version number, hash value, etc.), they can (5) compare the retrieved information with the information of their local copy of the software and the SBOM to verify its integrity. 

Let us revisit the attack where a malicious insider manipulates the version number of specific libraries and potentially even some source code as well. Changing the version number would result in looking up a different hash value from the repository that would not correspond to the one in the SBOM, which is computed from the actual software library during the SBOM generation stage. Similarly, any malicious changes to the software library itself would again result in a different hash value being included in the SBOM, which does not correspond to the value in the reference repository. The key point is that a malicious insider cannot tamper with the values in the reference repository nor with the computation of the hash value during SBOM generation. The latter is due to the fact that the SBOM generation happens after the stealthy attack and is not under control of the attacker. 

\subsection{Public repositories based on existing tools}
\begin{figure}[htbp]
    \centering
    \includegraphics[scale=0.4]{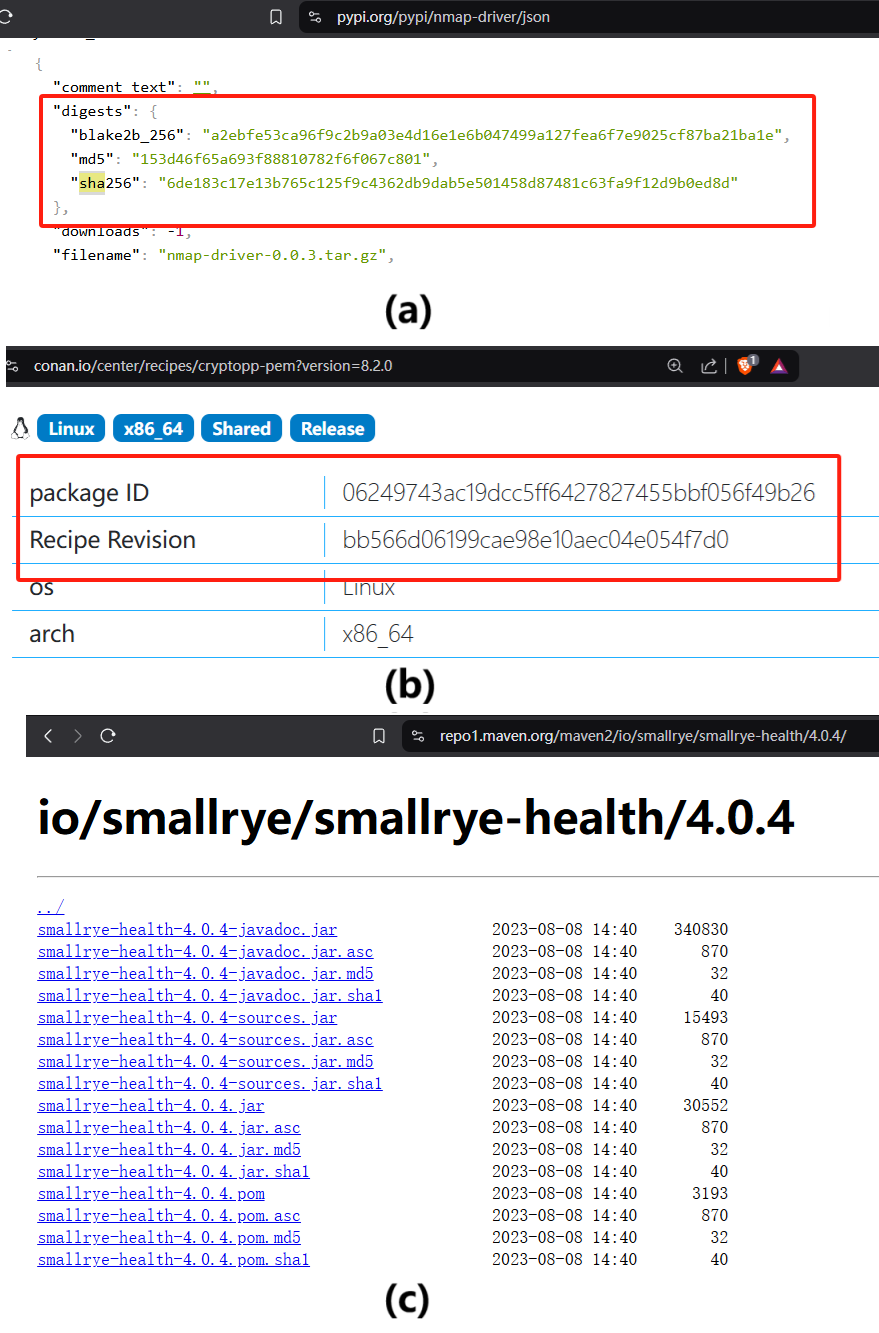}
    \caption{Hash values in code repositories for three (of eight) studied programming languages. Provided methods: (a) Python via .json files; (b) C/C++ via websites; (c) Java via separated files.}
    \label{fig:programming_language_hash}
\end{figure}

The key question is how to realize a reference repository for software libraries. Fortunately, various code repositories, such as PyPI and Maven, already exist for different programming languages. Therefore, they are a perfect candidate to serve as public reference repositories for SBOM verification by providing hash values for each library version. As shown in Table~\ref{tab:coderepo}, repositories are available for all eight languages studied. However, the format of hash value distribution varies across repositories, as illustrated in Fig.~\ref{fig:programming_language_hash}. For code repositories of Python (Fig.~\ref{fig:programming_language_hash}\textbf{.(a)}), C\#, JavaScript, and PHP, they provide a .json API, facilitating automation. For C/C++ (Fig.~\ref{fig:programming_language_hash}\textbf{.(b)}), the hash values are provided via the website. Finally, for Java (Fig.~\ref{fig:programming_language_hash}\textbf{.(c)}), they provide hash values through website-hosted files.

\begin{table}[htbp]
\caption{Dependency repositories for eight studied languages}
\begin{center}
\begin{tabular}{|c|c|}
    \hline
    \textbf{Programming Language} & \textbf{Public Repository}  \\ \hline
    Python & PyPI \\ \hline
    C &  Conan \\ \hline
    C++ & Conan \\ \hline
    Java & Maven \\ \hline
    C\# & Nuget \\ \hline
    JavaScript & npm \\ \hline
    PHP & Composer \\ \hline
    Rust & Cargo \\ \hline
\end{tabular}
\label{tab:coderepo}
\end{center}
\end{table}

In our proposed framework, the code repositories serve as secure software library repository. These public central repositories already provide hash values for most public libraries in their respective programming languages. The system remains dynamic, as hash values are automatically updated when developers release new versions. Therefore, one clear step forward in mitigating the malicious insider threat is to embed these repositories in the overall SBOM ecosystem.

\subsection{Proof-of-Concept Implementation}
To support our claims, we have implemented a proof-of-concept solution for secure SBOM generation and consumption for Python projects. It consists of both an SBOM verification module and an SBOM generator.

\begin{figure}
    \centering
    \includegraphics[width=\linewidth]{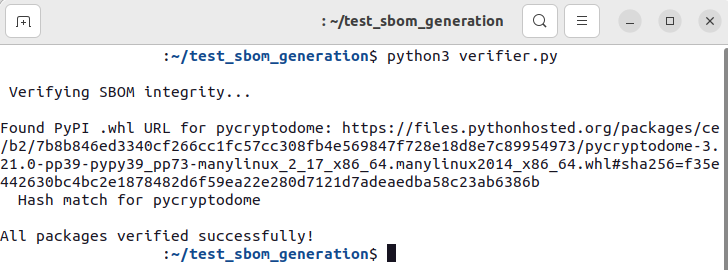}
    \caption{Screenshot of hash verification in our tool}
    \label{fig:hash_verification}
\end{figure}

Lets start with the former. Our verification tool systematically enumerates each dependency listed in the SBOM and then retrieves its corresponding SHA256 hash value directly from PyPI -- the public repository for Python. By comparing this retrieved hash value against the hash stored in the SBOM, our tool verifies the integrity and authenticity of each Python dependency used in the project, which in turn quickly identifies potential tampering or mismatches that could indicate a compromised SBOM file. Fig.~\ref{fig:hash_verification} shows a screenshot of this hash verification.

Our implementation iterates over each package defined within the SBOM, extracts the expected .whl filename and its stored cryptographic hash. For each dependency, the tool fetches the corresponding official hash from PyPI’s package registry by parsing the repository's HTML response. Next, our solution can detect any potential discrepancies by comparing the retrieved PyPI hash against the hash value stored in the SBOM. 

However, in the Python ecosystem, there are three main obstacles that need to be overcome. First, Pip, the package management system for the Python ecosystem, downloads the .whl (wheel) file temporarily into a cache location, installs it into your Python environment, and then typically discards the wheel file afterwards when a dependency is installed via pip3 install <package\_name>. In other words, the wheel file isn't permanently stored in your project directory or the current working directory under normal conditions. Second, the .whl file downloaded from PyPI has a fixed cryptographic hash (SHA256). This hash uniquely identifies the exact file contents at distribution. However, after installation, the wheel file is extracted, and files are placed into directories on the system (such as site-packages). This extraction process changes the structure and storage format, so the installed files no longer directly match the original wheel file’s hash. In other words, the hash you're verifying should always refer to the original, unmodified distribution artifact (the .whl file from PyPI), not the installed version of the package. Thus, hashes against the original downloaded wheel file from the official PyPI source should not be compared against the installed files. Therefore, each dependency's .whl file must be stored on the file system for hash comparison. 

\begin{figure}
    \centering
    \includegraphics[width=\linewidth]{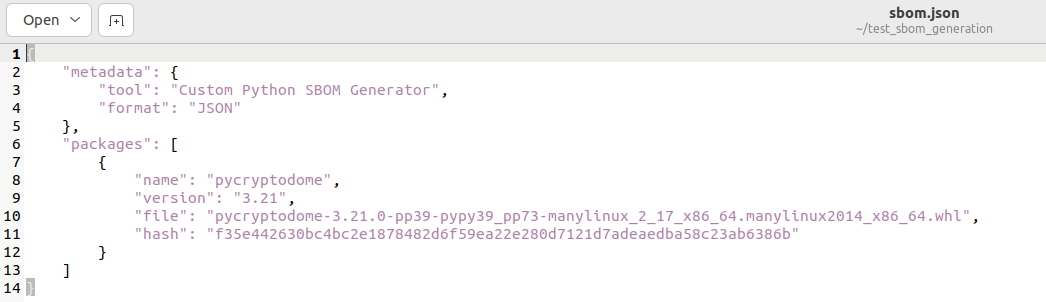}
    \caption{Screenshot of SBOM generation tool that computes hash values for dependencies}
    \label{fig:cyclonedxpy2}
\end{figure}

The last obstacle in securing SBOM dependencies against integrity attacks is that the officially recognized SBOM generator tool for Python does not generate a hash value for each dependency, as we mentioned in Section~\ref{malicious_developer_attack}. This makes SBOM hash comparison impossible. To overcome this gap, we also developed an SBOM generator tool for Python projects that generates a hash value for each dependency in the code base based on the .whl package, as seen in Fig.~\ref{fig:cyclonedxpy2}. 

One could argue that if an adversary, within our defined threat model, is capable of modifying the project's corresponding package manager and as such influence the SBOM generation process, why would an inside attacker then not tamper with the SBOM file, i.e., modify the dependency name and the corresponding hash values in the SBOM file? In this case, in-toto comes into play. We suggest to download all the dependencies' .whl file to the codebase and disallow the modification through in-toto. Therefore, even if there is an intentional or unintentional change in the package managers, our verifier will detect them as the hashes of the .whl files secured by in-toto will not match the ones from PyPI. 

In conclusion, Our SBOM verifier, combined with in-toto and our SBOM generator, ensures that any unauthorized changes or tampering by malicious insiders become detectable.

%% file: distributed.tex
\section{Decentralized approach}\label{decentralized}
In the previous section, we showed how trust in SBOMs can be established by relying on public repositories storing hash values of software libraries. However there might be cases where one cannot rely on this solution, for example when specific libraries are not included in any public repositories. Moreover one could argue one does not want to rely on a single source of trust to securely verify SBOMs. For these cases, we explore a decentralized solution based on blockchain. 

\subsection{Blockchain-based design}
\begin{figure}[htbp]
    \centering
    \includegraphics[scale=0.5]{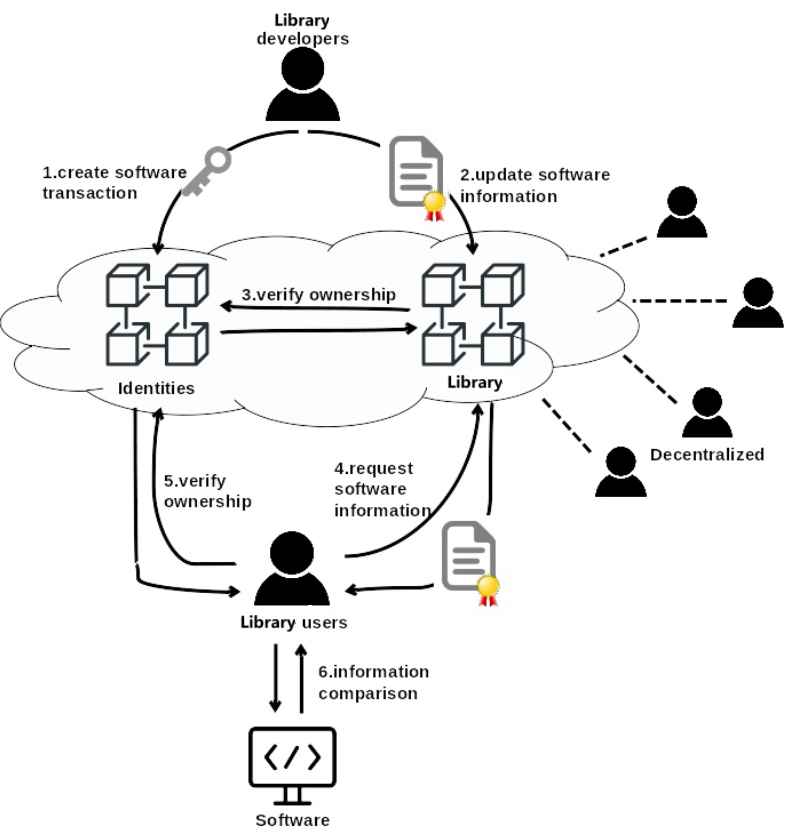}
    \caption{Decentralized framework for enhanced SBOM integrity protection}
    \label{fig:sbom_solution_framework}
\end{figure}

More specifically, we propose a distributed SBOM framework with integrity guarantees, as illustrated in Fig.~\ref{fig:sbom_solution_framework}. This framework is an extension of our proposed framework for secure SBOM generation, depicted in Fig.~\ref{fig:sbom_generation}, and also considers software library developers and users. Our extended framework contains two decentralized repositories (one for storing the identity information of the library developer/owner and another one for the information on the software library itself -- including the hash value). Since we now have moved to a decentralized setting, both software library developers and users can serve as decentralized contributors to these identity and library repositories. The identities repository serves as identity management and access control for the library repository. Developers can modify library information in the library repository only if they can prove ownership of this library. They do so by relying on the identities repository. On the other hand, the library repository stores records of software libraries, similarly as before (i.e., name and version of the software library, version number, hash value, etc.). 

\subsubsection{Block elements recorded in the two decentralized repositories}
The use of blockchain gives us integrity guaranties and supports transaction-based ownership control. The identities repository guarantees a developer's identity and their ownership of a specific library. Unlike a traditional blockchain that records cryptocurrency transactions, this identity blockchain will record software library transactions, such as the creation of a library (i.e., ownership) and the transfer of ownership. For instance, if developer A is the first to declare he created library L, the identities blockchain will log this, preventing other developers from claiming ownership of library L. Ownership can only change if developer A creates a transaction to transfer it to another developer, for example developer B. The library repository on the other hand stores the key details about the software library, such as its hash value, creation time, version number and other relevant identity information. Developers who own their library can upload this information to the blockchain as a reference for library users. The blockchain will deny the upload if a developer does not own the library. Additionally, library users can verify claims about a library -- in our setting this is the hash value of a specific version of the library -- by comparing it with the reference information stored on the blockchain. Each transaction in both the identities and library repositories is designed to include at least the following elements:

\textbf{\textit{For registering a library in the repository:}}
\textit{\begin{enumerate}
    \item Unique transaction id: A distinct identifier of the transaction.
    \item Timestamp: The time when the transaction was initiated.
    \item Owner public key: The public key of the library owner.
    \item Library name: Unique name of the library.
    \item Version: The version of the library.
    \item Hash value: Hash of this version of the library.
    \item Optional: Other, optional information.
    \item Signature: A signature created with the owner's private key.
\end{enumerate}}

\textbf{\textit{For transfer of ownership:}}
\textit{\begin{enumerate}
    \item Unique transaction id: A distinct identifier of the transaction.
    \item Timestamp: The time when the transaction was initiated.
    \item Input public key: The public key of the current library owner.
    \item Output public key: The public key of the new library owner. 
    \item Transactions: The information about the library for which ownership will be transferred.
    \item Signature: A signature created with the input's private key.
\end{enumerate}}

\subsubsection{Enhanced SBOM integrity protection}
As illustrated in Fig.~\ref{fig:sbom_solution_framework}, our framework is divided into two main parts: a registration phase and a verification phase. Lets now zoom in on the different steps: (1) To register a library, a developer must generate their public/private key pair and register their ownership of this library in the identities repository. This is only possible if the combination of library name and version number does not yet exist; otherwise, library ownership can only be transferred from the current owner (see further in this section). (2) Once ownership is confirmed, the developer uploads the library's details, such as its hash value, to the library repository. (3) The library repository then checks with the identities repository to confirm the developer's ownership. If confirmed, the information is successfully uploaded; if not, the upload fails. (4) If users want to verify the validity of the library, for example during the generation or consumption of an SBOM, users will make a query for the reference information to the library repository. Upon receiving the reference information, users check the transaction's signature using the owner's public key. (5) If the signature is verified, the identities repository is then consulted to confirm the owner's ownership of the library when the information was recorded in the library repository based on the owner's public key and the timestamp. (6) Once the identities repository confirms ownership, users can trust the reference information provided by the library repository and use this reference information in combination with an SBOM to validate the library they intend to use.

\subsubsection{Ownership transfer}
Ownership transfer occurs when a library owner decides to pass library ownership to another party. This is recorded in the identities repository. In our design, we consider a library name as a property, and transferring ownership means starting a transaction where the current owner assigns the library name to the new owner. Once recorded, the library repository can verify the new ownership details from the identities repository. 

\subsection{Feasibility analysis}
We argue that it is feasible to deploy a blockchain-based decentralized solution for storing and verifying reference information of software library libraries, including their hash values. Below we make an assessment of the storage, verification and registration speed.

\begin{table}[htbp]
\caption{The Size of Data Elements in the Identity and Library Repository}
\begin{center}
\begin{tabular}{|l|c|}
\hline
\textbf{Item}                 & \textbf{Size (byte)}\\ \hline
\textbf{Identity Repository}  &                     \\
ID (SHA-256)                  & 32                  \\
Timestamp                     & 8                   \\
Input Public Key (ECDSA-256)  & 32                  \\
Output Public Key (ECDSA-256) & 32                  \\
Transaction Content          & 128                 \\
Signature (ECDSA)             & 64                  \\ 
Total                         & 296                 \\ \hline
\textbf{Library Repository}  &                     \\
ID (SHA-256)                  & 32                  \\
Timestamp                     & 8                   \\
Owner Public Key (ECDSA-256)  & 32                  \\
Library Name                 & 32                  \\
Version Number                & 8                   \\
Hash Value                    & 32                  \\
Signature (ECDSA)             & 64                  \\ 
Total                         & 208                 \\ \hline
\end{tabular}
\label{tab:size_repository}
\end{center}
\end{table}

\subsubsection{Storage}
For our decentralized public repositories, the estimated size of each transaction is shown in Table~\ref{tab:size_repository}. The total size of a record in the identities and library repositories is 296 and 208 bytes, respectively. In 2023, Github reached 100 million developers and 284 million public repositories ~\cite{korkmaz2024github}. We assume that the average number of versions of each library is 10. Based on these numbers, we estimate the total storage size for a full node of the identities blockchain to be 27.57 GB, and similarly 550.15 GB for a full node in the library blockchain. It should be mentioned that this is most likely an over-estimation, as some public repositories on Github are documentation or resources instead of library, meaning that the actual total size will most likely be smaller. In comparison, the total size for a full node in Bitcoin was 562.61 GB on 21st May 2024 ~\cite{blockchainsize}. Therefore, we can conclude that the storage requirements for our framework (27.57 GB and 550.15 GB) are manageable. 

\subsubsection{Verification Speed}
Verification speed refers to how quickly a software library user can verify all the hashes of the library entries in an SBOM using blockchain. This process is similar to the ``check a transaction and the certificate'' speed described in ~\cite{wang2020blockchain}, which is 0.25 ms per verification. Based on this, we estimate a verification time of 0.5 ms and 1.5 ms for identity and library blocks, respectively, again based on 100 million users and 300 million public repositories. According to the npm official website in 2019 ~\cite{npmsummary2018}, the average modern web application has over 1000 dependencies. Thus, the estimated total verification time for a user is (0.5 + 1.5) $\times$ 1000 ms = 2 seconds, which includes verification of the hash value in the library blockchain and ownership verification in the identities blockchain. An SBOM verification process of 2 seconds is feasible, and further improvements could be achieved by caching frequently verified hash values of libraries. 

\subsubsection{Registration Speed}
Registration speed refers to how quickly blocks are created in the identities and library blockchain. To evaluate the registration speed, we collected the average confirmation time of Bitcoin over three years, from June 2021 to June 2024 ~\cite{blockavrtime}. The median of the average confirmation time was 40 minutes. We argue that a similar average confirmation time would be acceptable for the registration of software libraries in our decentralized repository system.

%% file: related_work.tex
\section{Related work}\label{Relatedwork}
A notable study has introduced a blockchain framework designed to improve SBOM sharing, featuring an authoritative structure and a tree-based certificate system. This framework capitalizes on the inherent integrity properties of blockchain technology, offering a potential solution for SBOM sharing challenges ~\cite{sbom_blockchain_2023}. However, issues remain with linking SBOMs to actual code, generating SBOMs with correct software hashes, and correctly consuming SBOMs. MITRE ATT\&CK proposed an end-to-end framework for software supply chain integrity, inspired by the SolarWinds incident. It emphasizes standardizing SBOMs, improving signature infrastructure, and enhancing automation, integration, and cryptographic traceability. Specifically, it advocates for strengthening the generation and distribution process of SBOMs, highlighting their critical importance ~\cite{mitre_sbom}. L.J. Camp et al. assess the risks associated with SBOMs and the immediate need for SBOM standardization. The authors argue that without widespread adoption of robust protocol security, SBOMs risk failing to meet their security promise ~\cite{camp2021sbom}. Although these efforts contribute to SBOM standardization and distribution, they largely overlook the challenges of SBOM generation and consumption, specifically establishing a reliable link between SBOMs and actual source code. Addressing these gaps is crucial to realize the full security benefits of SBOMs.

With the excellent performance of artificial intelligence (AI), people have started to apply AI in SBOM. B. Xia et al. propose "AIBOMs," an adaptation where traditional SBOMs, which list software components, are extended to suit AI systems' dynamic and evolving nature. AIBOMs specifically address the challenges posed by AI systems, such as reinforcement learning, which continuously evolves with new data, by creating a tailored BOM for these dynamic environments ~\cite{sbom_blockchain_2023}. Heeyeon Kim et al. developed a binary classifier to identify defective software prior to SBOM generation. By removing these defective software, the authors can produce an SBOM consisting solely of defect-free software ~\cite{kim2024deep}. These mechanisms add a security layer before SBOM generation, while they do not address the issue of linking SBOMs to the actual code.

For supply chain security, various tools have been developed, such as in-toto ~\cite{torres2019toto} and cosign from Sigstore ~\cite{newman2022sigstore}, both of which are promising candidates for improving SBOM integrity. In-toto is a framework designed to secure the software supply chain by ensuring the integrity of artifacts throughout the Software Development Lifecycle (SDLC). It provides a cryptographic mechanism to track each step in the SDLC, preventing unauthorized modifications. However, while in-toto can ensure that an SBOM file remains unaltered, it cannot verify whether the SBOM itself is correct, meaning it does not guarantee that the SBOM accurately represents the acutal software components. Similarly, cosign provides cryptographic signing and transparency logs for software artifacts, allowing developers to sign and verify SBOMs within the SBOM ecosystem. While this improves trust in provenance, cosign, like in-toto, ensures that an SBOM has not been tampered with but does not verify its correctness. The fundamental issue lies in the lack of a global hash comparison database that can validate whether the components listed in an SBOM match the actual software used. To the best of our knowledge, no global hash comparison framework exists for all programming languages. As a result, even with secure and robust SDLC security tools, we cannot fully guarantee the correctness of an SBOM. Our work demonstrates that attacks exploiting the missing link between SBOM files and actual software components pose a serious security threat. Furthermore, we propose a practical solution to address this gap, which is critical for ensuring the reliability and security of SBOMs in the software supply chain.

%% file: conclusion.tex
\section{Research Reproducibility}\label{Reproducibility}
We have released the software artifacts used in our research as well as the source code for our proof-of-concept SBOM verification and generation tool for Python. These can be found at the following (anonymous) GitHub repository \cite{repro}.

\section{Conclusion}\label{conclusion}
The use of SBOMs is widely advocated as a means to mitigate supply chain attacks, and there is an increase in legislation adopting this technology as a mandatory requirement when distributing software packages. In our work, we have performed an in-depth analysis of popular SBOM generation and consumption tools, and have identified several significant security weaknesses in both the tools and the processes in general. We have presented a malicious insider attack which allows to hide security vulnerabilities present in a software library (e.g., use of insecure software versions) in the SBOM generation process. This would result in a false sense of security and users of the software being unaware that patching is required. This stealthy tampering of the SBOM generation process is possible for six out of the seven most common programming languages. Our analysis of popular SBOM consumption tools showed that essential security checks, such as the verification of a digital signature on the SBOM, are missing. We have showed that the use of central repositories for storing reference hash values of software libraries, in combination with supply chain security techniques such as in-toto, is a viable mitigation strategy for the identified malicious insider attack. We have implemented a proof-of-concept solution for Python to demonstrate the feasibility of the proposed approach. In addition, we also explored the use of a decentralized solution as an alternative approach to improve the integrity verification of SBOMs.

%% file: acknowledgements.tex
\section*{Acknowledgments}
This work has been supported in part by CyberSecurity Research Flanders with reference number VR20192203, by the Energy Transition Fund of the FPS Economy of Belgium through the CYPRESS project, by the VLAIO COOCK program through the IIoT-SBOM project and by the Horizon Europe research and innovation program under grant agreement 101119747 TELEMETRY.